\documentstyle[12pt]{article}

\newcommand{\gv}{g_\perp^{(v)}}
\newcommand{\ga}{g_\perp^{(a)}}
\newcommand{\hs}{h_\parallel^{(s)}}

\newcommand{\htt}{h_\parallel^{(t)}}

\makeatletter
\def\slash#1{{\mathpalette\c@ncel{#1}}} 
\makeatother

\newcommand\beq{\begin{eqnarray}}
\newcommand\eeq{\end{eqnarray}}

\newcommand\la{\langle}
\newcommand\ra{\rangle}

\def\Dslash{\slash{\mkern-1mu D}}
\def\Delslash{\slash{\mkern-1mu \Delta}}
\def\Omslash{\slash{\mkern-1mu \Omega}}

\def\phat{\hat{p}}
\def\qhat{\hat{q}}

\begin{document}


\begin{titlepage}
\begin{flushright}
\begin{tabular}{l}
JUPD-9823\\
hep-ph/9805460
\end{tabular}
\end{flushright}
\vskip0.5cm
\begin{center}
  {\Large \bf 
             Renormalization of Chiral-Even Twist-3 
             Light-cone Wave Functions for Vector Mesons in QCD
  \\}

\vspace{1cm}
 {\sc Yuji~Koike}\,${}^a$, {\sc Naoki~Nishiyama}\,${}^b$ 
and {\sc Kazuhiro~Tanaka}\,${}^{c,}$\footnote{Supported 
in part by the Monbusho Grant-in-Aid for Scientific Research No. 09740215.}
\\[0.3cm]
\vspace*{0.1cm} ${}^a$ {\it Dept. of Physics, Niigata University,
Niigata 950--2181, Japan}
\\[0.3cm]
\vspace*{0.1cm} ${}^b$ {\it Graduate School of Science
and Technology, Niigata University,
Niigata 950--2181, Japan}
\\[0.3cm]
\vspace*{0.1cm} ${}^c$ {\it Dept. of Physics, Juntendo University,
Inba-gun, Chiba 270-1606, Japan}
\\[1cm]


  \vskip1.8cm
  {\large\bf Abstract:\\[10pt]} \parbox[t]{\textwidth}{ 
We present 
the one-loop anomalous dimension matrices for the chiral-even
twist-3 (nonsinglet) 
conformal operators, which govern the scale-dependence
of the vector meson light-cone wave functions
through the conformal expansion.
It is clarified that 
the constraints from
the charge-conjugation invariance 
and the chirality conservation allow only 
one independent anomalous dimension matrix for each conformal spin.

}

\end{center}

\vskip1cm


\noindent
[Keywords: Light-cone wave function, Twist three, Chiral-even,
Anomalous dimension matrix]

  \vskip1cm 
\end{titlepage}

\setcounter{equation}{0}

In a recent paper\,\cite{BBKT98}, 
we presented a systematic exploration 
on the twist-3 light-cone wave functions (distribution amplitudes)
of the vector mesons.
These wave functions are relevant for understanding 
preasymptotic corrections to various hard exclusive processes
producing vector mesons in the final state\,\cite{BLreport,CZreport}. 
According to that study,
two chiral-even wave functions $\gv(u)$, $\ga(u)$, and
two chiral-odd ones $\hs(u)$,  $\htt(u)$
constitute a complete set of the twist-3 quark-antiquark wave functions
($u$ denotes the light-cone momentum fraction 
carried by quark).  
The renormalization of these wave functions has also been studied 
based on the (approximate) conformal invariance in massless QCD\,\cite{O82}:  
The wave function can be 
decomposed into ``partial waves'' of definite
conformal spin, 
and the coefficients in this expansion 
are given as vacuum-to-meson 
matrix elements
of the conformal operators.  Conformal symmetry allows
renormalization mixing
only among the conformal operators with the same conformal spin
up to $O(\alpha_s^2)$ correction.  
It has also been shown in \cite{BBKT98} that 
the anomalous dimensions of those conformal operators 
can be expressed in terms of those for 
the nucleon's structure functions (parton distributions) 
{}for inclusive processes.
In particular,
the ``$\mu^{2}$-evolution''
of the chiral-odd wave functions
$\hs(u,\mu^2)$ and $\htt(u,\mu^2)$ has been solved in terms of
the anomalous dimension matrices for 
the nucleon's chiral-odd twist-3 
quark distribution functions $e(x,\mu^2)$  
and $h_L(x,\mu^2)$ \cite{KT95,BBKT96,KN97,BM97}.  
On the other hand, 
the anomalous dimension matrices for the chiral-even wave functions
have been only partially given in \cite{BBKT98} based on the known results 
{}for the nucleon's another chiral-even distribution,  
$g_{T}(x, \mu^{2})$\,\cite{g2,KYTU97}.

The purpose of this letter
is to complete the renormalization of the twist-3 conformal
operators relevant to chiral-even wave functions $\gv(u, \mu^{2})$ 
and $\ga(u, \mu^{2})$.
We give a convenient choice of the independent
operator basis, and clarify the symmetry constraints on their
renormalization mixing. 
We shall complete the relevant anomalous dimension 
matrices for all conformal spins.

Let us first introduce a twist-3 conformal operator basis
relevant for 
the renormalization of the nonsinglet chiral-even wave functions,
$\gv(u,\mu^2)$ and $\ga(u,\mu^2)$, 
and give precise formulation of our problem
referring to the results of \cite{BBKT98}. 
The QCD equations of motion allow us to reexpress
the quark-antiquark wave functions $\gv(u,\mu^2)$ and $\ga(u,\mu^2)$  
in terms of the twist-3 quark-antiquark-gluon wave functions.
This can be performed explicitly order by order in conformal expansion.
{}For the conformal partial wave of conformal spin $j= n+3/2$
($n= 2, 3, \cdots$), 
two sets of twist-3 quark-antiquark-gluon conformal operators,
$\{R^{+}_{n,k}\}$ and
$\{R^{-}_{n,k}\}$,
come into play:
\beq
& &R^{\pm}_{n,k} = 
\Theta^V_{k,n-k-2} \pm \Theta^V_{n-k-2,k} 
\mp \Theta^A_{k,n-k-2} + \Theta^A_{n-k-2,k} \qquad (k=0,1,\ldots,n-2),
\label{eq1.1}\\
& &\Theta^V_{k,n-k-2}(0)
\equiv \bar{\psi}(0)d^{n-k-2}
\Delslash
g\widetilde{G}^{\perp \lambda}(0)\Delta_\lambda
d^{k}\psi(0)+ ({\rm total\ derivatives}),
\label{eq1.2}\\
& &\Theta^A_{k,n-k-2}(0)
\equiv \bar{\psi}(0)d^{n-k-2}
\Delslash gG^{\perp \lambda}(0)\Delta_\lambda
d^{k}i\gamma_5\psi(0)
+ ({\rm total\ derivatives}),
\label{eq1.3}
\eeq
where we introduced an auxiliary 
light-like vector $\Delta^\mu$ 
with $\Delta^{+} = \Delta^{\perp} = 0$,
which guarantees the symmetrization of the $n$ Lorentz indices
and the subtraction of the trace terms, 
and $d\equiv i\Delta\cdot D$ with $D_\mu =\partial_\mu -igA_\mu$
the covariant derivative.  
{}For the purpose of renormalization,  
it suffices to consider the flavor-diagonal operators
as given above. 
The dual gluon field strength  
is defined as $\widetilde{G}_{\mu\nu}= 
{1\over 2}\epsilon_{\mu\nu\lambda\sigma}
G^{\lambda\sigma}$ where $\epsilon_{\mu\nu\lambda\sigma}$ is the
totally anti-symmetric tensor
with $\epsilon_{0123}=1$.  (We follow the convention
of \cite{BBKT98} in this paper).
In (\ref{eq1.2}), `total derivatives'
stands for the terms consisting of
$$\left(\Delta\cdot \partial\right)^{n-r}
\left[\bar{\psi}d^{r-j-2}
\Delslash
g\widetilde{G}^{\perp \lambda}\Delta_\lambda
d^{j}\psi\right],\ \  (\ 2 \le r \le n-1,\ 0\le j \le r-2),$$ 
and similarly for
those in (\ref{eq1.3}).
These total derivative terms are organized so that the operators
(\ref{eq1.2}) and (\ref{eq1.3}) form irreducible representation
of conformal spin $j$ of the so-called 
collinear conformal group\cite{O82}.
The complete expression for $\{\Theta^{V,A}_{k,n-k-2}\}$
is given in terms of
the Appell polynomials,
but it is irrelevant here.
The suffix $k$ of (\ref{eq1.1})-(\ref{eq1.3}) labels 
many independent conformal operators
having the same conformal spin,
illustrating that three-particle representations of conformal group
are degenerate;
the number of independent operators increases with the spin.
Conformal symmetry does not allow renormalization mixing
in the leading logarithmic order
between the operators with different $n$;
it does allow, however, mixing between the operators with different $k$
{}for the same $n$.
Here we note that there exists a constraint from symmetry:
$\{R_{n,k}^+ \}$ and $\{R_{n,k}^- \}$
have a definite charge-conjugation 
parity $(-1)^{n}$ and $(-1)^{n+1}$, respectively,
and do not mix with each other under renormalization.  
If we define the anomalous dimension matrix for 
$\{R_{n,k}^\pm\}$ as $\Gamma_n^\pm$, 
the renormalization group (RG) equation for $\{R^\pm_{n,k}\}$
is given by
\beq
\mu\frac{d}{d\mu}R^\pm_{n,k}(0; \mu^2) = -{\alpha_s \over 2\pi}
\sum_{l=0}^{n-2}\left(\Gamma^\pm_n\right)_{k,l}R^\pm_{n,l}(0; \mu^2),
\label{eq1.RG}
\eeq
where $\mu\frac{d}{d\mu}=
\mu{\partial \over \partial \mu} 
+ \beta(g){\partial\over \partial g}$ with $\beta(g)$ the $\beta$-function.
This equation is solved to give 
the scale-dependence of $R_{n,k}^\pm$ as
\beq
R_{n,k}^\pm(0; Q^2) =\sum_{l=0}^{n-2} \left[
\left(\frac{\alpha_s(Q^2)}{\alpha_s(\mu^2)}\right)^{\Gamma^\pm_n/b}
\right]_{k,l}
R_{n,l}^\pm(0; \mu^2),
\label{eq1.4}
\eeq
where 
$b={11\over 3}N_c -{2\over 3}N_f$.
The mixing matrix $\Gamma_{n}^{\pm}$ is subject to our computation.
The nucleon's forward matrix 
elements of $R^{+}_{n,k}$ with even $n$ is associated with the $n$-th moment
of the 
transverse spin structure function $g_2(x,Q^2)$
in the deep inelastic scattering.  In this connection
the renormalization of $R^{+}_{n,k}$
has been studied by several different approaches 
\cite{g2,KYTU97} for even $n$,
while the result for odd $n$ has not been considered seriously.
On the other hand, there have been no places in inclusive processes where 
$R^-_{n,k}$ appear, and their renormalization
has never been discussed before.

\vskip0.5cm

In order to obtain the anomalous dimension matrix for 
the conformal operator basis $\{ R^\pm_{n,k} \}$,
it is convenient to work with their forward matrix elements:
We imbed $\{ R^\pm_{n,k} \}$ into the three-point function as
\begin{equation}
\int d^{4}x d^{4}y d^{4} z e^{ipx-iqy+i(q-p)z}
\langle 0 | {\rm T} R^{\pm}_{n,k} (0) \psi(x) \bar{\psi}(y)
A_{\mu}^c(z) |0 \rangle. 
\label{eq:n1}
\end{equation}
Then,
the total derivative terms in
(\ref{eq1.2}) and (\ref{eq1.3}) drop out 
and the calculation becomes greatly
simplified.  For later convenience, we introduce
two sets of operators, which replace $R_{n,k}^{\pm}$ in (\ref{eq:n1}):
\begin{eqnarray}
W^{(a) \sigma}_{n,k} &=& 
U^{\sigma}_{(+)k} + U^{\sigma}_{(-)n-k-2},
\label{eq:wa} \\
W^{(v) \sigma}_{n,k} &=& 
U^{\sigma}_{(+)k} - U^{\sigma}_{(-)n-k-2},
\label{eq2.Wpm}
\end{eqnarray}
where
$k=0,1,\ldots,n-2$, and
\begin{equation}
U^{\sigma}_{(\pm)k} =
\bar{\psi}d^{n-k-2}\Delslash
\left(
g\widetilde{G}^{\sigma\lambda}\Delta_\lambda 
\pm
i\gamma_5
g{G}^{\sigma\lambda}\Delta_\lambda 
\right)
d^k\psi.
\label{eq:n2}
\end{equation}
Note that  
$W_{n,n-k-2}^{(a) \sigma=\perp}$ and $W_{n,k}^{(v) \sigma=\perp}$
agree with $R_{n,k}^{+}$ and $R_{n,k}^{-}$, respectively, 
if one ignores the total derivatives from the latter. 
{}For the renormalization, 
we follow the method of \cite{KT95,KN97,KYTU97}:  We
calculate the one-loop corrections for (\ref{eq:n1}).
To renormalize the composite operators involving 
massless fields consistently,
we keep the quark and gluon external lines off-shell.
We then find that the following twist-3 operators also participate:  
$\{W_{n,k}^{(a)\sigma}\}$ mix with\footnote{
The operators $W_{n,F}^{(a)\sigma}$,
$\frac{1}{2(n+1)}W_{n,k}^{(a)\sigma}$, and $W_{n,E}^{(a)\sigma}$
coincide with a set of operators considered in \cite{KYTU97}
{}for the massless quark limit.}
\beq
W_{n,F}^{(a)\sigma}
&=& \frac{1}{n+1}\left[n
\bar{\psi}\gamma_5\gamma^\sigma d^n \psi 
-\sum_{k=1}^n \bar{\psi}\gamma_5 \Delslash d^{k-1}iD^\sigma
d^{n-k}\psi \right],
\label{eq2.F+}\\
W^{(a)\sigma}_{n,E}
&=& {n\over 2(n+1)}\left[
\bar{\psi}i\Dslash\gamma_5\gamma^\sigma\Delslash
d^{n-1}\psi 
+\bar{\psi}\gamma_5\gamma^\sigma\Delslash d^{n-1}
i\Dslash\psi
\right],
\label{eq2.EOM+}
\eeq
while 
$\{W_{n,k}^{(v)\sigma}\}$ mix with
\beq
W_{n,F}^{(v) \sigma}
&=&\frac{1}{n+1}
i \epsilon^\sigma_{\  \mu \nu \tau} \ell^{\mu}\Delta^{\nu} \left[n
\bar{\psi}\gamma^\tau d^n \psi
-\sum_{k=1}^n \bar{\psi}\Delslash d^{k-1}iD^\tau
d^{n-k}\psi
\right],
\label{eq2.F-}\\
W^{(v)\sigma}_{n,E}
&=& {n\over 2(n+1)}
i \epsilon^\sigma_{\  \mu \nu \tau} \ell^{\mu}\Delta^{\nu} \left[
-\bar{\psi}i\Dslash \gamma^\tau\Delslash
d^{n-1}\psi 
+\bar{\psi}\gamma^\tau\Delslash d^{n-1}
i\Dslash \psi
\right],
\label{eq2.EOM-}
\eeq
where we introduced another light-like vector $\ell^{\mu}$
with 
$\ell^{-}=\ell^{\perp}=0$
and $\ell\cdot \Delta = 1$.
In the following, we assume $\ell^\mu$ and $\Delta^\mu$
have mass dimensions $+1$ and $-1$, respectively. 
The forward matrix elements of the quark bilinear operators
$W^{(a)\sigma}_{n,F}$ and $W^{(v)\sigma}_{n,F}$ coincide with
those of the twist-3 quark-antiquark conformal operators
of spin $j=n+3/2$ for $\ga(u, \mu^{2})$
and $\gv(u, \mu^{2})$.
It is instructive to note that they
originate from the
light-cone nonlocal operators 
$\bar{\psi}(0)\gamma_5\gamma^\sigma[0,\Delta]\psi(\Delta)$
and $\bar{\psi}(0)\gamma^\sigma[0,\Delta]\psi(\Delta)$,
where $[0,\Delta]\equiv {\rm exp}\left\{-ig \int_0^1\,dt 
\Delta_\mu A^\mu(t\Delta)
\right\}$ is the gauge link operator:
Taylor expansion of these nonlocal operators 
at small quark-antiquark separations
yields
$\bar{\psi}(\gamma_5)\gamma^\sigma iD^{\{\mu_1}\cdots iD^{\mu_n\}}\psi$,
and
the subtraction of the twist-2 components, which are totally symmetric
and traceless among $\{\sigma,\mu_1,\cdots,\mu_n\}$,
leads to $W^{(a, v)\sigma}_{n,F}$
corresponding to
antisymmetric pairs of indices
$\sigma$ and $\mu_{i}$.
$W^{(a, v)\sigma}_{n,E}$ are the so-called equation-of-motion (EOM) 
operators, which contribute as nonzero operators 
to the off-shell Green's function.
As it is well known, 
the above twist-3 operators are not independent, 
but obey the following relation:
\begin{equation}
W^{(a, v)\sigma}_{n,F}= 
{1\over 2(n+1)}\sum_{k=0}^{n-2}(k+1)W^{(a,v)\sigma}_{n,k}
+ W^{(a,v)\sigma}_{n,E}. 
\label{eq2.+relation}
\end{equation}
Therefore, we can conveniently choose (\ref{eq:wa}) and (\ref{eq2.EOM+})
((\ref{eq2.Wpm}) and (\ref{eq2.EOM-}))
as an independent basis for renormalization.
A physical matrix element of $W^{(a,v)\sigma}_{n,E}$
vanishes, and does not affect the final results of the 
scale-dependence.
However, it has to be kept in the 
intermediate step of the calculation to work out the 
renormalization\,\cite{KT95,KN97,KYTU97}.

Here we comment on 
the places where these operators 
appear in inclusive processes involving nucleons.  
The nucleon matrix element $\la PS|W_{n,F}^{(a)\perp}(Q^{2})|PS\ra$
gives
the twist-3 part of 
the $n$-th moment
$\int_{-1}^1 dx x^n g_T(x, Q^{2})$
with 
$g_T(x, Q^{2})$ the nucleon's transverse parton distribution. 
($|PS\ra$ is the nucleon state with momentum $P^{\mu}$ and
spin vector $S^{\mu}$). 
{}For even $n$, this moment is associated with 
$\int_{0}^{1}dx x^{n}g_{2}(x, Q^{2})$
measurable in the deep inelastic scattering, while both even and odd 
moments are relevant 
in the polarized Drell-Yan process (see e.g. \cite{JJ92}).  
In order to identify the role of $W^{(v)\sigma}_{n,F}$ in 
inclusive processes,
we recall that it is generated from the nonlocal
light-cone operator $\bar{\psi}(0)\gamma^\sigma[0,\Delta]\psi(\Delta)$ and
consider the following decomposition 
of the nucleon matrix element, 
\begin{equation}
\int {d\lambda\over 2\pi}e^{i\lambda x}\la PS |\bar{\psi}(0)\gamma_\mu
[0, \lambda\Delta] \psi(\lambda \Delta)|PS\ra = 2 \left[ f_1(x)\ell_\mu 
+ \epsilon_{\mu\nu\rho\sigma}
\ell^\nu \Delta^\rho S_\perp^\sigma f_T(x) + M^2 \Delta_\mu f_4(x)  \right],
\label{eq2.ft}
\end{equation}
where 
$P^{\mu}$ and $S^{\mu}$
are decomposed as $P^\mu = \ell^\mu + {M^2\over 2}\Delta^\mu$ and
$S^\mu=(S\cdot \Delta)\ell^\mu + (S\cdot \ell)\Delta^\mu
+S_\perp^\mu$ with the nucleon mass
$M$ and $P^{2}=- S^{2}=M^{2}$.
$f_{1}(x)$ and $f_{4}(x)$ are the spin-independent parton distribution 
{}functions of twist-2 and 
twist-4, respectively, as discussed in \cite{JJ92}.  
The new function 
$f_{T}(x)$ is of twist-3\cite{BMT98}.  
By comparing both sides of (\ref{eq2.ft}),
it is easy to see that 
$\int_{-1}^1dx x^n f_T(x, Q^{2})$ is associated with 
$\la PS|W^{(v)\perp}_{n,F}(Q^{2})|PS\ra$.
However, the existence of $f_{T}(x)$ violates time reversal
invariance, and therefore $\la PS|W^{(v)\perp}_{n,F}|PS\ra
=\la PS|W^{(v)\perp}_{n,k}|PS\ra=0$.
(The nonforward matrix elements
like $\la 0|R^{\pm}_{n,k} |P\ra$
relevant to meson wave functions  
are not constrained to vanish
by time reversal invariance.)

In order to work out the
renormalization of $W^{(a,v)\sigma}_{n,k}$, 
we clarify the
symmetry constraints on the renormalization mixing. We find:

\begin{enumerate}

\item[(i)] Two sets $\{U_{(+)k}^{\sigma}\}$ and 
$\{U^{\sigma}_{(-)k}\}$ in (\ref{eq:n2}) do not
mix with each other.  
To see this, we recall $\sigma=\perp$ and 
introduce right- and left-handed circular polarization vectors
{}for the gluon as
$R=(0, 1, i, 0)/\sqrt{2}$ and $L=(0, 1, -i, 0)/\sqrt{2}$, respectively. 
Then we find 
$
U_{(+)k}^{\sigma = R}= -2i\bar{\psi}_L 
d^{n-k-2}\Delslash G^{R\nu}\Delta_\nu  d^k\psi_L$ and  
$
U_{(-)k}^{\sigma=R}= -2i\bar{\psi}_R 
d^{n-k-2}\Delslash G^{R\nu}\Delta_\nu  d^k\psi_R$ where
$\psi_{R,L} = (1\pm\gamma_5)\psi/2$.  
We also obtain similar expressions
with $R \leftrightarrow L$ for $U^{L}_{(+)k}$ and $U^{L}_{(-)k}$.
These results  
explain no mixing between
$\{U^{\sigma}_{(+)k}\}$ and $\{U^{\sigma}_{(-)k}\}$, 
since the perturbative quark-gluon coupling 
preserves chirality of each quark
line.

\item[(ii)] The anomalous dimension matrices for 
$\{U^{\sigma}_{(+)k}\}$ and $\{U^{\sigma}_{(-)k}\}$ are identical: 
{}From (i), the RG equation for $\{U^{\sigma}_{(+)k}\}$ 
can be written with its anomalous dimension matrix $\Gamma_n$ as 
\beq
\mu\frac{d}{d\mu}
U^{\sigma}_{(+)k} = -{\alpha_s\over 2\pi}
\sum_{l=0}^{n-2}\left(\Gamma_n\right)_{k,l}U^{\sigma}_{(+)l}, 
\label{RGtheta+}
\eeq
where the terms due to the EOM operators are ignored 
assuming that we take a certain physical matrix element. 
By applying the charge conjugation ${\cal C}$
on both sides, one obtains
\beq
\mu \frac{d}{d\mu}
U^{\sigma}_{(-)n-k-2} = -{\alpha_s\over 2\pi}
\sum_{l=0}^{n-2}\left(\Gamma_n\right)_{k,l}U^{\sigma}_{(-)n-l-2}, 
\label{RGtheta-}
\eeq
where we used 
${\cal C}U^{\sigma}_{(+)k}{\cal C}^{-1} = (-1)^{n}U^{\sigma}_{(-)n-k-2}$. 
Equations (\ref{RGtheta+}) and (\ref{RGtheta-})
explain the statement. 

\end{enumerate}
{}From (\ref{RGtheta+}) and (\ref{RGtheta-}), 
one obtains the RG equation for $W^{(a,v)\sigma}_{n,k}$ as
\beq
\mu \frac{d}{d\mu}W^{(a,v)\sigma}_{n,k} = -{\alpha_s\over 2\pi}
\sum_{l=0}^{n-2}\left(\Gamma_n\right)_{k,l}W^{(a,v)\sigma}_{n,l}. 
\label{RGW}
\eeq
By comparing this result with forward matrix element of (\ref{eq1.RG}),
we can identify
the anomalous dimension matrix $\Gamma_{n}^{\pm}$ as
\begin{equation}
\left(\Gamma^+_n\right)_{k,l} = \left(\Gamma_{n}\right)_{n-k-2, n-l-2};
\;\;\;\;\;
\left(\Gamma^-_n\right)_{k,l} = \left(\Gamma_{n}\right)_{k, l}.
\label{eq:nn}
\end{equation}
We also note that
we can use whichever 
convenient set of the operators
$\{ W^{(a,v)\sigma}_{n,k}\}$,
$\{ U^{\sigma}_{(\pm)k}\}$,  
in order to obtain $\Gamma_{n}$ (see (\ref{RGtheta+})-(\ref{RGW})).
These features of the anomalous dimensions
are in parallel with a 
{}familiar example of the twist-2 distributions
$f_1$ and $g_1$
defined from the nonlocal operators
$\bar{\psi}(0)\gamma^\sigma[0,\Delta]\psi(\Delta)$ and
$\bar{\psi}(0)\gamma_5\gamma^\sigma[0,\Delta]\psi(\Delta)$:
Their $n$-th moment is associated with the local operators   
$\bar{\psi}d^n\Delslash\psi
=\bar{\psi}_Rd^n\Delslash\psi_R + \bar{\psi}_L d^n\Delslash\psi_L$ and 
$\bar{\psi}d^n\Delslash\gamma_5\psi
=\bar{\psi}_Rd^n\Delslash\psi_R - \bar{\psi}_L d^n\Delslash
\psi_L$, respectively,
and similar argument as above 
explains that 
the nonsinglet parts of $f_1$ and $g_1$ have the same 
anomalous dimensions. 

\vskip0.5cm

Explicit calculation of the mixing matrix $\Gamma_{n}$ is yet 
desirable for general (even and odd) $n$.
As noted above, one has to 
keep the EOM operators as nonzero operators during the course of the 
renormalization.
We choose $\{W^{(a,v)\sigma}_{n,k},
W^{(a,v)\sigma}_{n,E}\}$  
as a basis of the renormalization. 
(For completeness, we parallely discuss two sets of the operators.) 
We imbed these operators into the three-point function (\ref{eq:n1}).
The relevant one-loop diagrams are the same as those considered in
\cite{KYTU97}.
We adopt the Feynman gauge, and obtain
the mixing matrix $\Gamma_{n}$ 
in the MS scheme of the dimensional regularization.
One technical difficulty in the computation is 
the complicated 
mixing with many gauge noninvariant EOM operators.
To avoid this,
we introduce a vector $\Omega^\mu$ with the 
condition $\Omega\cdot \Delta=0$, 
and contract the external gluon line of (\ref{eq:n1})
with this vector\,\cite{KT95}.    
Under this condition, 
the vertices 
{}for $W^{(a,v)\sigma}_{n,k}$ and $W^{(a,v) \sigma}_{n,E}$ 
can be written as  
\beq
{\cal W}^{(a,v)\sigma}_{n,k} &=&
\pm g(\qhat-\phat)
\left( \phat^k\qhat^{n-k-2}\gamma_5\gamma^\sigma
\Omslash\Delslash \mp \qhat^k\phat^{n-k-2}\gamma_5 
\Omslash\gamma^\sigma\Delslash \right)t^c,
\label{eq2.base1}\\
{\cal W}^{(a,v) \sigma}_{n,E} &=&
\mp {n\over 2(n+1)}g\left( \phat^{n-1}\gamma_5\gamma^\sigma
\Omslash\Delslash \pm \qhat^{n-1}\gamma_5 
\Omslash\gamma^\sigma\Delslash \right)t^c,
\label{eq2.base2}
\eeq
where $t^c$ is the color matrix normalized as 
${\rm tr}(t^{c}t^{c'})= \frac{1}{2}\delta^{cc'}$,
and we introduced the notation $\hat{p}=p\cdot\Delta$, etc. 
It is now straightforward to obtain
the mixing matrix $\Gamma_{n}$.
The result reads
\beq
\left(\Gamma_n\right)_{kl} &=&
-C_G{ (l+1)(l+2) \over (k+1)(k+2)(k-l)}\nonumber\\
& &+(2C_F-C_G)\left[ 
(-1)^{k+l+1}{ {}_{n-1}C_l \over {}_{n-1}C_k }{n+k-l \over n(k-l)}
+(-1)^{l+1}{ {}_{k}C_l \over k+2} \right],\qquad (k>l)
\label{eq2.gamma1}\\
\left(\Gamma_n\right)_{kk} &=&
C_G\left(
{-1\over n-k-1} + {1\over n-k} + {1\over n-k+1} + {1\over k+2}
+S_{n-k-1} + S_{k+1} \right)\nonumber\\
& &+ (2C_F - C_G)\left[ 
{-1\over n} + { 2(-1)^{n-k} \over (n-k-1)(n-k)(n-k+1) }
+ { (-1)^{k+1} \over k+2} \right]\nonumber\\
& &+ C_F \left( 2S_{k+1} + 2S_{n-k-1} - 3 \right),
\label{eq2.gamma2}\\
\left(\Gamma_n\right)_{kl} &=&
-C_G { (n-l)(n-l+1) \over (n-k)(n-k+1)(l-k)}
+(2C_F-C_G)\left[ 
(-1)^{k+l+1}{ {}_{n-1}C_{l+1} \over {}_{n-1}C_{k+1} }
{ n+l-k \over n(l-k) }\right.\nonumber\\ 
& &\left.\qquad\qquad+ { 2(-1)^{n-l} {}_{n-k-1}C_{l-k}
\over  (n-k-1)(n-k)(n-k+1) } \right],\qquad (l>k)
\label{eq2.gamma3}
\eeq
where ${}_{n}C_{k} = n!/[k!(n-k)!]$, 
$S_n=\sum_{j=1}^n\left(1/j\right)$, and
$C_{F}=(N_{c}^{2}-1)/2N_{c}$, $C_{G} = N_{c}$ for $N_{c}$ color.
This result combined with (\ref{eq:nn}) gives $\Gamma_{n}^{\pm}$.
The result for $\Gamma^-_n$ is wholly new.
{}For $\Gamma^+_n$, the result is now completely applicable
also for the cases with odd $n$
which has not received much attention.\footnote{
$\Gamma_{2}^{-}$, $\Gamma_{3}^{+}$, and $\Gamma_{4}^{-}$
give Eqs. (4.66), (4.67) and (4.68) of \cite{BBKT98}.}   
{}For example, 
if one replaces the term $(-1)^l/(n-l)$ proportional
to $2C_F - C_G$ in Eq. (16) of \cite{KYTU97} by 
$(-1)^{n-l+1}/(n-l)$ in their convention,
their result in Eqs.~(15)--(17) becomes consistent with 
the above (\ref{eq2.gamma1})--(\ref{eq2.gamma3}) for all $n$. 

We make contact with the work of Balitsky and Braun\cite{BB88},
who also considered the renormalization of similar
quark-antiquark-gluon operators.  
They introduced the 
nonlocal light-cone operators
\beq
S^{\pm}_\sigma(\alpha,\beta,\gamma) = g\bar{\psi}(\alpha \Delta)
[\alpha \Delta, \beta \Delta]
\left( G_{\sigma\nu}(\beta \Delta) \pm \widetilde{G}_{\sigma\nu}
(\beta \Delta)i\gamma_5
\right)
\Delta^\nu \Delslash [\beta \Delta, \gamma \Delta] 
\psi (\gamma \Delta),
\label{eq:non1}
\eeq
and computed the one-loop evolution kernels
in the coordinate space representation.
They found by explicit calculation
that $S_{\sigma}^{+}(\alpha, \beta, \gamma)$
and $S_{\sigma}^{-}(\alpha, \beta, \gamma)$
are renormalized without mixing with each other.
The reason for this phenomenon
is made clear by examining the 
$\sigma = R, L$ components of (\ref{eq:non1})
similarly to (i) above.
A direct relation between our results and those of \cite{BB88}
can be provided by expanding the operators (\ref{eq:non1})
in powers of $\Delta_{\mu}$ and taking the forward matrix element: 
Expansion of $S_{\sigma}^{\pm}(\alpha, \beta, \gamma)$
generates a series of the local operators with increasing spin.
In particular, the $n$-th term generates
$\Delta^{\nu}\ell^{\lambda}
\epsilon_{\sigma \nu \lambda \mu}U^{\mu}_{(\pm)k}$
with $k = 0, 1, \ldots, n-2$, up to the twist-4 corrections.
In this connection, we mention the following point: 
The kernels for $S_{\sigma}^{+}(\alpha, \beta, \gamma)$
and $S_{\sigma}^{-}(\alpha, \beta, \gamma)$,
given by Eq.(6.2) of \cite{BB88},
have rather different forms,
although we have shown that the
anomalous dimension matrices
{}for the corresponding two sets of local operators,
$\{U^{\sigma}_{(+)k}\}$ and $\{U^{\sigma}_{(-)k}\}$,
are identical with each other
(see (\ref{RGtheta+}) and (\ref{RGtheta-})).
In fact, the equivalence of these two kernels
can be easily proved
by applying the parity transformation combined with
the time-reversal transformation\footnote{
This transformation is more convenient for this purpose
than the charge-conjugation transformation used 
in (\ref{RGtheta+}) and (\ref{RGtheta-}), since it
connects $S_{\sigma}^{+}(\alpha, \beta, \gamma)$
with $S_{\sigma}^{-}(\alpha, \beta, \gamma)$
keeping the condition $\alpha > \gamma$, which
is assumed in Eq.(6.2) of \cite{BB88}.}
to the corresponding evolution equations,
and therefore the results of \cite{BB88} are consistent with
ours.

We finally recall in our notation that the anomalous dimension matrix in
(\ref{eq2.gamma1})--(\ref{eq2.gamma3})
satisfies the relation in the large $N_c$ limit, i.e. $C_F\to N_c/2$
in (\ref{eq2.gamma1})--(\ref{eq2.gamma3})\,\cite{ABH91}:
\beq
& &\sum_{k=0}^{n-2}(k+1)\left(\Gamma_n\right)_{k,l} = (l+1)\gamma_n,
\label{eq:nnnn}
\eeq
where $\gamma_n$ is the lowest eigenvalue of the anomalous dimension
matrix $\Gamma_n$ in this limit, and is given by 
\beq
\gamma_n=2N_c\left( \psi(n+1) +\gamma_E -{1\over 4} + {1\over 2(n+1)} 
\right).
\label{eq:evenlargeNcAD}
\eeq
Here $\psi(n+1) = \sum_{k=1}^{n}1/k - \gamma_{E}$ is the digamma function
and $\gamma_{E}$ is the Euler constant.
This leads to a
simple evolution equation in the large $N_{c}$ limit
(see (\ref{eq1.4}), (\ref{RGW}), and (\ref{eq:nn})): 
\beq
{\cal O}_n(Q^{2})= 
\left( \frac{\alpha_{s}(Q^{2})}{\alpha_{s}(\mu^{2})}\right)^{\gamma_n/b}
{\cal O}_{n}(\mu^{2}),
\label{eq:nnn}
\eeq
where ${\cal O}_{n}(\mu^{2}) = 
\sum_{k=0}^{n-2}(n-k-1)R_{n,k}^{+}(0; \mu^{2}),
\sum_{k=0}^{n-2}(k+1)R_{n,k}^{-}(0; \mu^{2})$,
and $W_{n,F}^{(a,v)\sigma}(\mu^{2})$.
Similar relations (\ref{eq:nnnn}) and (\ref{eq:nnn}) hold
also for the limit of large conformal spin
with 
$\gamma_{n}\rightarrow \gamma_{n} + (4C_{F} -2N_{c})(\ln n +\gamma_{E}
- 3/4)$ to $O(\ln n/n)$ accuracy.
As a result of (\ref{eq:nnn}), the vector meson wave functions
$\ga(u,\mu^2)$, $\gv(u,\mu^2)$ as well as the nucleon parton 
distribution $g_{T}(x, \mu^{2})$ obey simple
DGLAP-type evolution in the two limits, $N_{c} \rightarrow \infty$
and $n \rightarrow \infty$\cite{BBKT98,ABH91}.

To summarize, 
we have completed the renormalization of chiral-even twist-three
nonsinglet wave functions for the vector mesons. 
The result is presented in the form of the anomalous dimension matrices
{}for the corresponding 
conformal operators $\{R^\pm_{n,k}\}$ ($k=0,1,\cdots,n-2$)
{}for all conformal spins.
Although $\{R^+_{n,k}\}$ and $\{R^-_{n,k}\}$ are the independent
operators, their anomalous dimension matrices coincide completely 
due to the charge-conjugation invariance and 
the chirality conservation.

\end{document}